# Adaptive Embedding Pattern for Grayscale-Invariance Reversible Data Hiding

Erdun Gao[1], Zhibin Pan[1,2,*], Xinyi Gao[1]

1. School of Electronic and Information Engineering, Xi'an Jiaotong University, Xi'an 710049, P. R. China.

2. Key Laboratory of Spectral Imaging Technology, Chinese Academy of Sciences, Xi'an 710119, China

**Abstract**

In traditional reversible data hiding (RDH) methods, researchers pay attention to enlarge the embedding capacity (EC) and to reduce the embedding distortion (ED). Recently, a completely novel RDH algorithm was developed to embed secret data into color image without changing the corresponding grayscale [1], which largely expands the applications of RDH. In [1], for color image, channel R and channel B are exploited to carry secret information, channel G is adjusted for balancing the modifications of channel R and channel B to keep the invariance of grayscale. However, we found that the embedding performance (EP) of that method is still unsatisfied and could be further enhanced.

To improve the EP, an adaptive embedding pattern is introduced to enhance the competence of algorithm for selectively embedding different bits of secret data into pixels according to context information. Moreover, a novel two-level predictor is designed by uniting two normal predictors for reducing the ED for embedding more bits. Experimental results demonstrate that, compared to the previous method, our scheme could significantly enhance the image fidelity while keeping the grayscale invariant.

## I. INTRODUCTION

Copyright protection and content authentication of digital multimedia are always hot topics

which draw more and more interest of researchers. In some certain situations, secret information should be transmitted through cover image without attracting the attackers' interest. To this end, reversible data hiding (RDH) was proposed to embed secret data into cover image while restore both the cover image and secret data completely. That is to say, RDH causes no damage and totally ensures the integrity of the cover image. With this good characteristics, RDH is particularly applied in some sensitive fields, such as content authentication [2], data coloring in the cloud [3], medical image processing [4].

Until nowadays, many efficient methods [5–17] have been designed for applications on gray image. On one hand, data hiding technique should reduce the distortions brought to the cover media, since slight modification and high visual quality could help to distract the attackers' interest. On the other hand, larger embedding capacity (EC) brings the advantage that less cover media would be needed to transmit a fixed amount of secret information. Recently, with color images popularizing [18–22], researchers used to exploit the correlation among three channels (R, G, B) for further minimizing the distortion based on cooperating with classical methods on grayscale. In fact, when switching to the field of color image, the main aim is still the same as that of gray image. Therefore, unfortunately, image structure is inevitably damaged by the embedding procedures, which may further affect the following-up utilizations of marked image, such as feature extraction [23,24], key point matching [25]. For expanding the further applications of RDH, attention should be paid to guarantee the structure characteristics of marked image even though the cover image could be restored in the receiving end.

To address the aforementioned problem, creatively, Hou et al. [1] compromised that the gray version of cover image could be invariant to at least guarantee the gradient information, which is actually essential and sufficient for the above mentioned assignments. Moreover, gray version could obviously reduce the computation complexity. By ensuring this, after data hiding, marked image could be further exploited for other applications. In [1], a novel algorithm is designed to achieve the above requirement. In the cover image, red channel and blue channel are set to carry secret information, the green channel is adjusted to eliminate the modifications of red and blue channels. For enhancing the embedding performance (EP), a polynomial predictor based on gray version is also designed to improve the predicting accuracy. Finally, benefitting from the immutability of the gray version, the secret information could be well extracted after predicting and recovering the red and blue channel values.

In our consideration, keeping features invariant is really an important issue but the performance of distortion controlling is still unsatisfied in [1]. By carefully investigating, we found that in [1], only one bit of secret data could be embedded into one unit (including three channels) with that all three channels need to be modified, which really results in the low embedding efficiency. Firstly, to evaluate the distortion of unit, we design an indicator named unit embedding distortion (UED), by which we could find that the distortions of channel B and channel G are nearly fixed in every unit. Based on this observation, therefore, in our proposed method, we attempt to optimize the embedding pattern by adaptively embedding one or two bits of secret data into channel R according to context information. In other words, for pixels located in smooth regions, they should be further exploited instead of being processed totally the same as those pixels located in normal regions. Furthermore, for better fitting with our new embedding pattern, we cooperatively design a novel two-level predictor, in which the predicting and embedding procedures are operated alternately. This idea comes from that the second embedding would cause worse image quality reduction than the first embedding. Hence, optimizing the second embedding operation is more important to keep higher image quality. By the above two innovations, our method achieves a lower embedding distortion and could also keep the grayscale invariant.

The remainder of this paper is organized as follows. In Section II, Hou et al.'s work, which is the backbone of our proposed method, is introduced. Section III presents the proposed method in details including the adaptive embedding pattern and the two-level predictor. Performance evaluation and comparisons with [1] are demonstrated in Section IV. Finally, we conclude the paper in Section V.

## II. HOU ET AL.'S WORK

### A. Polynomial predictor

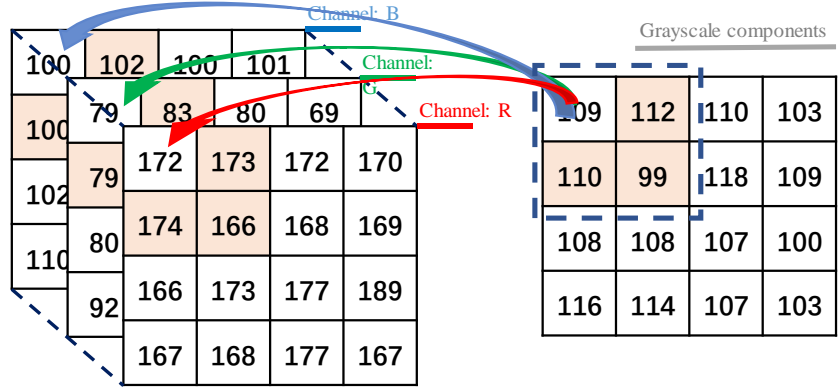

Fig. 1. Illustration of polynomial predictor.

Just like the traditional method demonstrating [26], sharper prediction error distribution could ensure superior embedding performance, which is more suitable for data embedding. From this fact, a polynomial predictor is designed in [1], which holds the view that pixel in each channel could be predicted by the corresponding the value of grayscale with a non-linear polynomial formulate:

$$X_{i,j}^{p} = [1, gr_{i,j}, gr_{i,j}^{2}] \begin{bmatrix} a \\ b \\ c \end{bmatrix} \quad (1)$$

where $X_{i,j}^{P}$ is the predicted value and $gr_{i,j}$ represents the value of grayscale in the position of $(i, j)$. And, $\beta = [a, b, c]^{T}$ is the parameter set, in which the three parameters need to be optimized adaptively. For pixel of each position, three neighboring pixels are set as inputs to find the fittest $\beta$ represented as $\beta^{o}$ by minimizing the following formula (taking blue channel as example):

$$\beta^{o} = \arg\min_{\beta} \|K - I\beta\|^{2} \quad (2)$$

where $I = \begin{bmatrix} 1 & gr_{i+1,j} & gr_{i+1,j}^{2} \\ 1 & gr_{i,j+1} & gr_{i,j+1}^{2} \\ 1 & gr_{i+1,j+1} & gr_{i+1,j+1}^{2} \end{bmatrix}$, $K = [b_{i+1,j}, b_{i,j+1}, b_{i+1,j+1}]^{T}$ and the optimal $\beta$ could be calculated by:

$$\beta^{o} = (I^{T}I)^{-1}I^{T}K \quad (3)$$

The idea of median-edge detector (MED) is also introduced to design this predictor (The details could be found in [1]). Finally, error could be obtained by:

$$e = b - b^p \quad (4)$$

Like difference expansion (DE) [5], the obtained predicted errors are expanded to carry secret data.

B. *Choice of embedding position*

When trying to transform color image into grayscale, the most popular method is to use the following function:

$$\begin{aligned} gr &= f_{c2g}(r,g,b) \\ f_{c2g} &\to round(0.299r + 0.587g + 0.114b) \end{aligned} \quad (5)$$

where *round()* means to output the nearest integer, and *r, g, b* represent the pixel values of channel R, channel G, channel B, respectively. And *gr* is the value of grayscale.

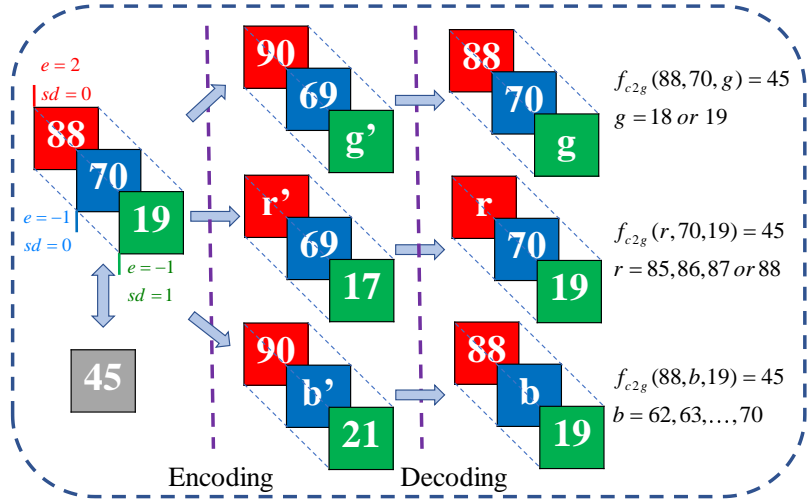

Fig. 2. Choices of different error correcting channel.

To ensure grayscale invariant, as shown in Fig. 2, two channels are chosen for data embedding and the other channel is modified for keeping the grayscale invariant. However, as shown in Eq. (5), all coefficients are decimals, which means that there may be more than one value that own the same grayscale when recovering the original pixel. For ensuring the reversibility, we need to add error correcting bit (ECB) for distinguishing the original value of pixel. In the three different choices, it is best to choose channel G as the adjusted channel for only one bit is enough for meeting this requirement.

Consequently, as shown in Fig. 3, channel R is chosen for embedding secret bit, while channel B is embedded with the ECB of the last unit as shown in Fig. 4, channel G carries no secret data but is modified for keeping the grayscale invariant.

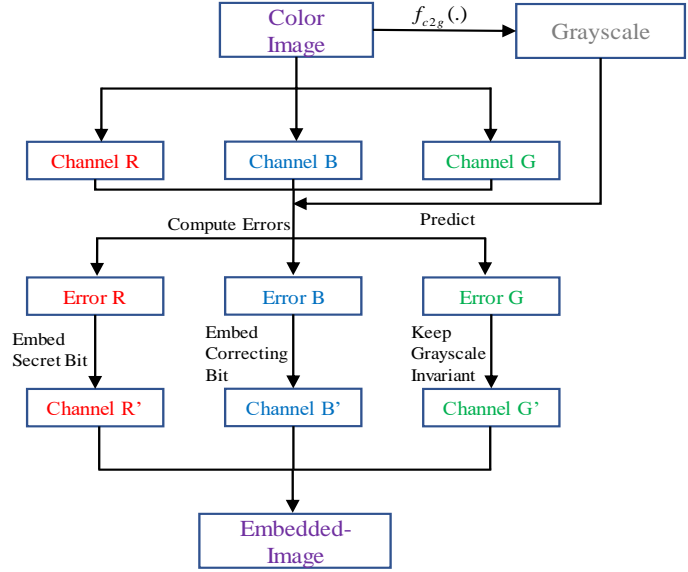

Fig. 3. Embedding procedure of Hou et al.'s method.

## III. PROPOSED METHOD

This section introduces our novel method including the adaptive embedding pattern and the two-level predictor. Moreover, we give the details of mechanism to verify the superiority of our two-level predictor. The procedures of the algorithm are also presented for the embedding and extraction. For better illustrating, a concrete example is presented. Finally, in our method, some auxiliary information is processed, which encodes the thresholds and the length of location map.

### A. Adaptive embedding pattern

As it has been illustrated before, for three channels, only one bit of secret data could be embedded in [1]. Obviously, it is a waste of the embedding position and a limit to the performance. Here, for clearly explaining, we take three channels (R, G, B) as one unit. Therefore, the unit embedding distortion (UED) of data embedding could be calculated as Eq. (6).

$$UED = d(r) + d(b) + d(g) \qquad (6)$$

where *d* is the distortion function of Euclidian distance (L-2 norm) for measuring the modification of pixel value. In [1], channel B is utilized for transmitting auxiliary information and channel G is adjusted for keeping the grayscale invariant, which, obviously, cannot be altered under this framework. In this view, another straightforward idea for reducing the UED is to embed more secret data into one unit.

For measuring the performance of RDH algorithms, researchers used to apply peak-signal-noise-ratio (PSNR) as the criterion. Note that, PSNR is not linearly related with the modification of pixels, but square. To this end, we need taking care of not modifying one pixel too much. Concretely speaking, it is important that not try to embed secret data into pixels which cause large prediction errors.

Starting from the above considerations, we propose to embed two bits of data rather than just one bit in channel R if current unit is located in smooth region, which means that we could get high predicting accuracy. And then, the UED could be reduced as given in the following equation:

$$UED = \frac{d_2(r) + d(b) + d(g)}{2} \quad (7)$$

where $d_2$ represents the total distortions of embedding twice. The new embedding pattern is shown in Fig. 4. It could be seen that when a pixel unit is classified as smooth region, two secret bits are embedded into the red channel. Meanwhile, the afore coming error correcting bit (ECB) is embedded into the blue channel. Finally, for keeping the grayscale invariant, the green channel is adjusted to balance the modification of red and blue channels. When a pixel unit is denoted as normal region, one and only difference is that we embed one bit of secret data into the red channel. For the other channels, all are the same.

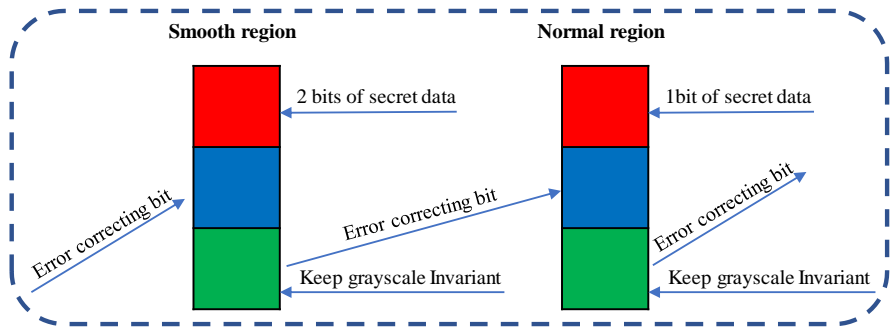

Fig. 4. Embedding pattern of two different units.

Here, it may be hard to directly understand that this operation could significantly benefit the performance of distortion-capacity. A simple example would be given to better demonstrate this advantage in the following subsection *D*.

*B. Two-level predictor*

Actually, the simplest way [26] for embedding two bits of secret data *sd* into one pixel is to expand the prediction error (*PE*) twice (we take $PE \leq 0$ phase as an example),

$$\begin{aligned} P_{wm2} &= P_{original} + PE_2 - sd_2 \\ &= P_{original} + 2(PE_1 - sd_1) - sd_2 \end{aligned} \quad (8)$$

Where $P_{original}$ and $P_{wm2}$ represent the original value and modified value of current pixel, respectively. However, the relationship between the distortion and the modification is not linear. In other words, the second embedding ($sd_2$) may cause larger distortion than the first embedding ($sd_1$). To address this issue, unlike the conventional way, we design a novel predictor to reduce the second embedding distortion by re-predicting the pixel value after the first embedding.

Firstly, we introduce two classical predictors named median-edge detector (MED) [27] and accurate gradient selective prediction (AGSP) [28], respectively.

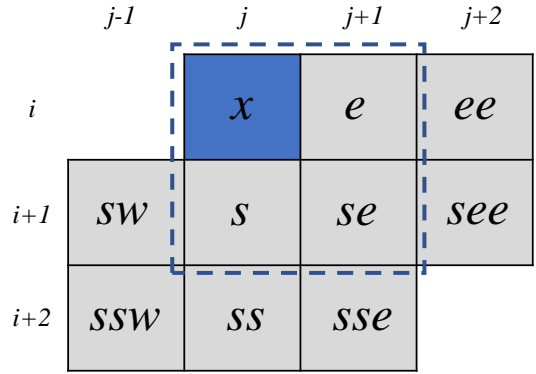

Fig. 5. Diagram of neighboring pixels of current pixel *x*.

Both MED and AGSP are simple yet effective predictors, which utilize the gradient information to predict the value of current pixel *x*. According to Fig. 5, the predicted value $P_{MED}$ of *x* given by MED is calculated as follows:

$$P_{MED} = \begin{cases} \min(e,s), & \text{if } se \geq \max(s,e) \\ \max(e,s), & \text{if } se \leq \min(s,e) \\ s+e-se, & \text{otherwise.} \end{cases} \quad (9)$$

The AGSP operates on the nine neighboring pixels of the current pixel $x$ as reference shown in Fig. 5, The estimated gradient and the predicted value $P_{AGSP}$ of $x$ is calculated as follows:

Horizontal direction:

$$D_H = (2|e-ee| + 2|s-se| + 2|s-sw| + |ss-sse| \\ + |ss-ssw| + |se-see|)/9 + 1 \quad (10)$$

Vertical direction:

$$D_V = (2|e-se| + 2|s-ss| + |sw-ssw| + |ee-see| \\ + |se-sse|)/7 + 1 \quad (11)$$

+45-degree direction:

$$D_{+45} = (2|e-s| + 2|s-ssw| + |se-ss| + |ee-se|)/6 + 1 \quad (12)$$

-45-degree direction:

$$D_{-45} = (2|e-see| + 2|s-sse| + + |sw-ss|)/5 + 1 \quad (13)$$

For each direction in the above four, casual pixels $s, e, sw, se$ are chosen, respectively. In these four directions, we seek the two smallest gradients as $D_{\min 1}, D_{\min 2}$ and the corresponding casual pixels as $C_{\min 1}, C_{\min 2}$. Finally, the predicted value is calculated as follows:

$$P_{AGSP} = \frac{D_{\min 1} \times C_{\min 2} + D_{\min 2} \times C_{\min 1}}{D_{\min 1} + D_{\min 2}} \quad (14)$$

In our method, we propose to conduct the embedding procedure and predicting procedure alternately. Firstly, we take the average value of MED and AGSP as the first predicted result of $P_{original}$. And then, the first secret bit is embedded as follows:

$$\begin{aligned} P_1 &= (P_{MED} + P_{AGSP})/2, \\ PE_1 &= P_{original} - P_1, \\ P_{wm1} &= P_{original} + PE_1 - sd_1 \end{aligned} \quad (15)$$

$P_{wm1}$ is the modified value after embedding the first secret bit $sd_1$. And then, the second step is as follows:

$$\begin{aligned}
P_{\min} &= \min\{P_{MED}, P_{AGSP}\}, \\
P_{\max} &= \max\{P_{MED}, P_{AGSP}\}, \\
P_2 &= \begin{cases} P_{\min}, & \text{if } P_{wm1} \leq P_{\min} \\ P_{\max}, & \text{if } P_{wm1} \geq P_{\max} \\ NULL, & \text{otherwise} \end{cases}, \\
PE_2 &= P_{wm1} - P_2, \\
P_{wm2} &= P_{wm1} + PE_2 - sd_2
\end{aligned} \tag{16}$$

In the above formulas, $P_1$ and $P_2$ represent the predicted values of $P_{original}$ and $P_{wm1}$. $PE_1$ and $PE_2$ represent the prediction errors of the two steps, respectively. $P_{wm1}$ and $P_{wm2}$ are the modified pixel values after embedding, $sd_1$ and $sd_2$ are the secret bit.

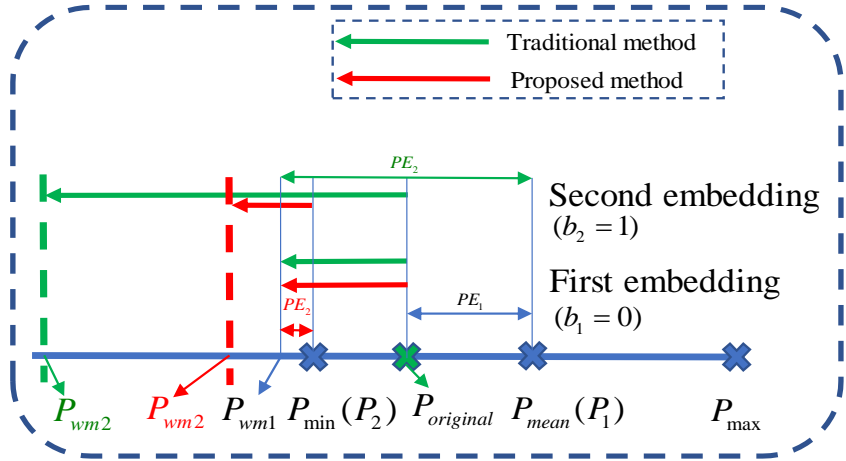

Fig. 6. Different embedding mechanisms of traditional method and the proposed method.

From Fig. 6, we could see that when trying to embed the first secret bit $sd_1$, the given two methods are totally the same. However, when trying to embed the second secret data $sd_2$, traditional method simply expands the error which contains the first predicted error and embedding data. Differently, our predictor could generate a smaller prediction error for the second embedding by re-predicting, which helps to reduce the distortion.

C. *Procedures of the proposed method*

In our proposed method, a location map *LM* is generated to avoid the overflow and underflow, and $\Delta$ is set to describe the complexity of the region where the current pixel locates in.

$T_1$ and $T_2$ $(T_1 \leq T_2)$ are two thresholds to classify the smooth, normal and complex regions. $\Delta$ is defined as the variance of current pixel and its four neighboring pixels ($Xv$ represents the grayscale): $\{Xv_{i,j}, Xv_{i-1,j}, Xv_{i+1,j}, Xv_{i,j-1}, Xv_{i,j+1}\}$.

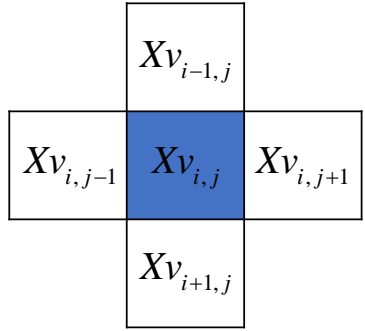

Fig. 7 Four neighboring pixels for calculating complexity.

Note that the grayscale always keeps invariant. Therefore, the complexity information could be ensured the same at both the encoding and decoding ends. Finally, Fig.8 shows the example of embedding procedure of one unit.

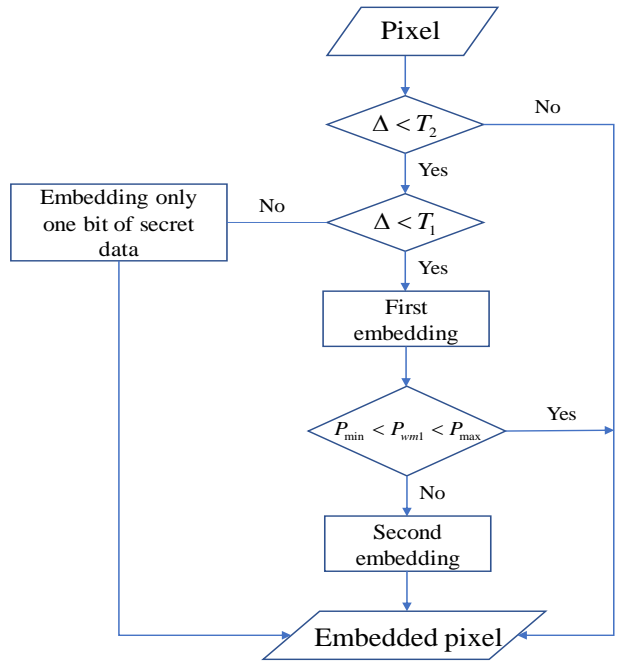

Fig. 8 Embedding example of one unit.

*D. A simple example of one unit*

In this subsection, we give an instance of the embedding operations of Hou et al.' method, Li et al.'s [26] method and our proposed method. And, we compare the unit embedding distortion (UED) of these three methods to verify the superiority of our proposed method.

Before the comparisons, we firstly introduce the measurement of evaluating the performance of algorithms. Mostly, peak-signal-noise-ratio (PSNR) is applied as the indicator, which is defined as

$$PSNR = 10\log_{10}(\frac{255^2}{MSE}) \quad (17)$$

and MSE is calculated by

$$MSE = \frac{1}{M \times N \times C}\sum_{k=1}^{C}\sum_{i=1}^{M}\sum_{j=1}^{N}(p_v(i,j,k) - p'_v(i,j,k))^2 \quad (18)$$

where M, N and C represent the width, height and channel of the cover image, respectively. Generally, for color image, C is equal to 3 including red channel, blue channel and green channel. $p_v(i,j,k)$ and $p'_v(i,j,k)$ represent the pixel values at the position $(i,j,k)$ of the cover image and marked image.

**Embedding example of smooth region (Hou et al.'s)**

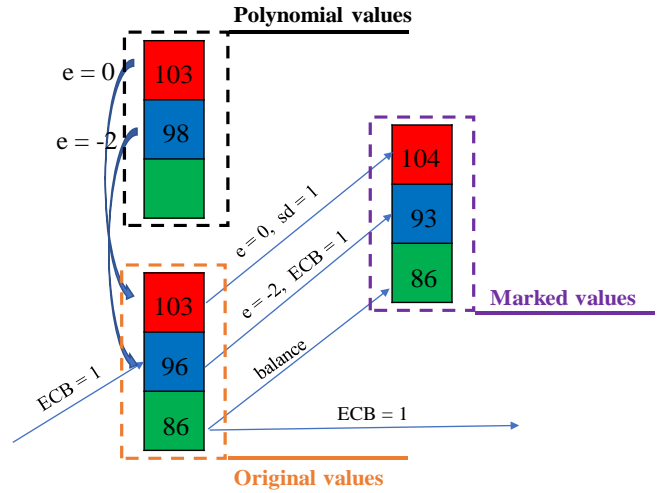

Fig. 9 Embedding example of one unit in smooth region (Hou et al.'s).

As is shown in Fig. 9, in this unit, the orginal values of three channel (R, G, B) are 103, 86 and 96. By the polynomial predictor, the predicting values of channel R and channel B are calculated as 103 and 98. The secret data (*sd*) is binary code which is 1 here, and error correcting bit (ECB) of the previous unit is 1. For the channel R, prediction error is 103-103=0 and the *sd* is embedded. So the marked value is 104. For the channel B, prediction error is 96-98=-2 and the ECB is embedded.

So the marked value is 93. For the channel G, predicting procedure is not needed, because it is adjusted for balancing the modifications of channel R and channel B to keep the grayscale invariant.

For indicating the distortion of one unit, we use the unit embedding distortion (UED) as defined by Eq. (6). Actually, due to PSNR is related to the size and channel number of the cover image, which can not utilized for one unit. Thus, here, we keep the form of square to evaluate the UED. In this example, we could get that the UED is

$$UED_{Hou} = (104-103)^2 + (93-96)^2 + (86-86)^2 = 10 \tag{19}$$

**Embedding example of smooth region (Li et al.'s)**

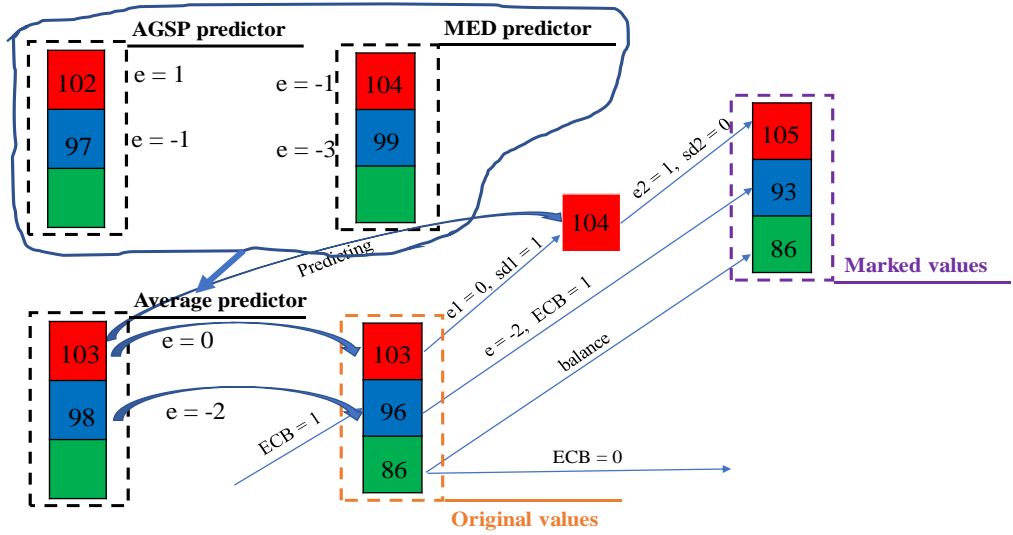

Fig. 10 Embedding example of one unit in small region (Li et al.'s).

Fig. 10 shows the example of Li et al.'s method. For clearly comparing the embedding mechanism with our proposed method, we also take the average value of median-edge detector (MED) predictor and accurate gradient selective prediction (AGSP) predictor as the first predicting result. Different from Hou et al.' method, a second embedding operation is effected to the channel R, where the predicting value is just the same as the first result. Therefore, for Li et al.'s method, we could obtain the UED is

$$UED_{Li} = \frac{(105-103)^2 + (93-96)^2 + (86-86)^2}{2} = 6.5 \tag{20}$$

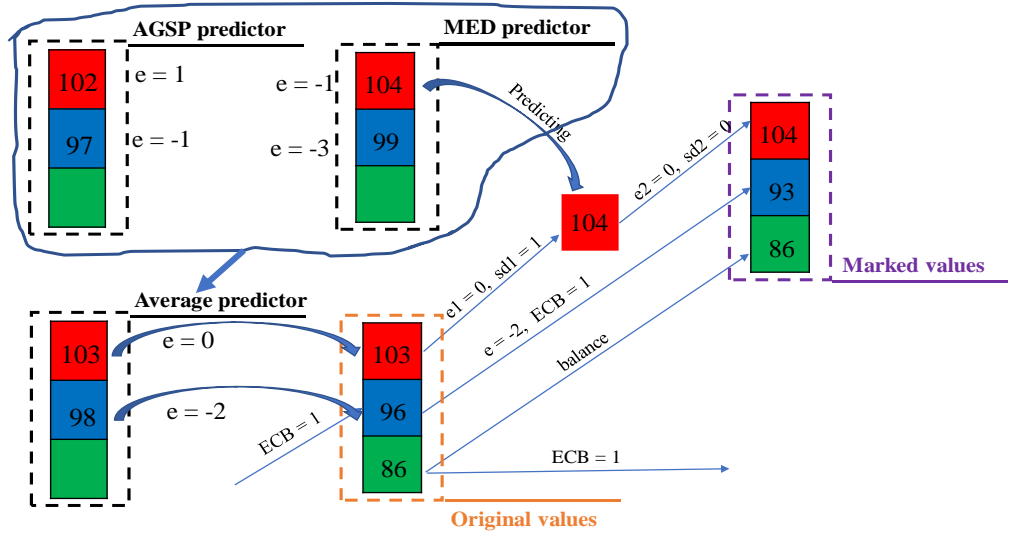

Fig. 11 Embedding example of one unit in smooth region (Ours).

As shown in Fig. 11, in our proposed method, we design a novel two-level predictor. For the second embedding, we do not use the same result as that in Li et al.'s method. Instead, we use one of the predicting results of MED predictor and AGSP predictor according to the value of the first marked value. In this example, the first embedding value is 104, so MED predictor is utilized to get the second predicting result. Here, we could calculate the UED of our proposed method is

$$UED_{Ours} = \frac{(104-103)^2 + (93-96)^2 + (86-86)^2}{2} = 5 \qquad (21)$$

From the given example, we could find that comparing to the fixed embedding pattern, adaptive embedding pattern is more efficient. Even utilizing the ordinary embedding method proposed by Li et al. , the UED could be reduced by $(10-6.5)/10 = 35\%$. This could obviously verify the superoitiry of the adaptive embedding pattern.

Comparing to Li et al.' method, our proposed method could optimize the second embedding procedure by getting a new prediciton result. Finally, our method could reduce the UED by $(10-5)/10 = 50\%$ and $(6.5-5)/6.5 = 23.1\%$ respecively when comparing to the other methods.

## Embedding example of normal region (Ours)

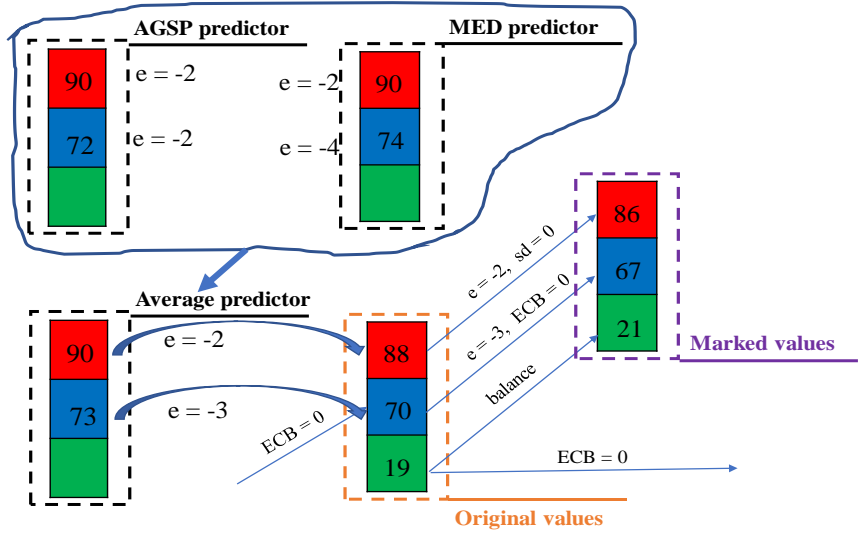

Fig. 12 Embedding example of one unit in the normal region (Ours).

In Fig. 12, we could see a simple embedding example of one unit in the normal region. For this case, we embed only one bit of secret data just like the embedding pattern of [1]. Because, for this unit, we could calculate that the UED of the first embedding operation is $(86-88)^2 + (67-70)^2 + (21-19)^2 = 17$, which is too high. If we insist embedding one more bit in this case, the distortion would be horriblely large.

## Extraction example of smooth region (Ours)

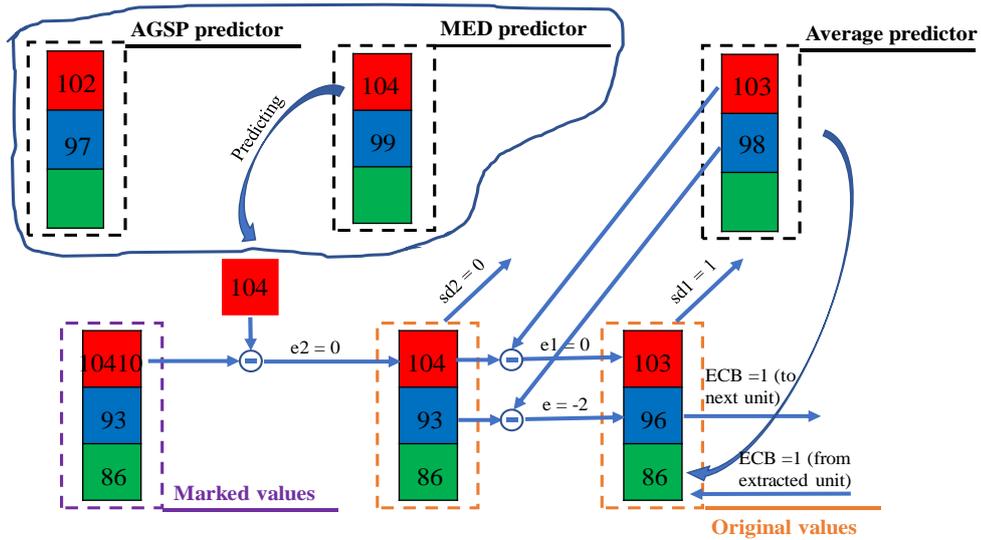

Fig. 13 Extraction example of one unit in the smooth region (Ours).

Here, we provide the extraction example of the case where unit is in smooth region. Firstly, we find that the marked values of this unit are 104, 93 and 86. Meanwhile, we could also obtain the predicting results of MED predictor and AGSP predictor for they are only related to the grayscale which keep invariant all the time. Secondly, for the channel R, according to the value of MED predictor and marked value of channel R, the second secret bit and the transitional pixel could be extracted, which are 0 and 104 here. And based on the values of average predictor and transitional pixel, the first secret bit and the original pixel value could be recovered. For the channel B, only the average predictor and marked value are utilized for once, from which we could easily get the original value and the ECB for the next unit. For channel G, no prediction procedure is implemented. From the ECB of last unit and the grayscale, the original value could be calculated as

$$g_{original} = round((gr - 0.299 r_{marked} - 0.144 b_{marked})/0.587) \qquad (22)$$

*E. Auxiliary information*

Notice that, pixels are unable to be shifted when the marked value exceed the interval [0, 255], this will cause overflow/underflow. Here, we construct a location map ($LM$) to record these points. If the unit has such kinds of pixels, we set $LM(i)=1$ and $LM(i)=0$ in other cases, where $i$ represents the index of unit. Next, location map is compressed into a shorter bit stream denoted $SLM$ by arithmetic coding. The bit-length of $SLM$ is $L_{clm}$. Additionally, in proposed method, besides the location map, some parameters should be transmitted as auxiliary information.

(1) The block complexity thresholds $T_1$ (8 bits), $T_2$ (8 bits);

(2) The end position $K_{end}$ ($\lceil \log_2 N \rceil$ bits);

(3) Length of the compressed location map $L_{clm}$ ($\lceil \log_2 N \rceil$ bits);

Thus, the size of auxiliary information is $(16 + 2\lceil \log_2 N \rceil)/3$. In order to ensure reversibility, LSBs of the first $(16 + 2\lceil \log_2 N \rceil)/3$ units are recorded $S_{LSB}$. Then, replace these LSBs by our auxiliary information and add $S_{LSB}$ to the secret data. Here, we need to claim that, not all the grayscale of marked image is the same as cover image. The first $(16 + 2\lceil \log_2 N \rceil)/3$ units could not be avoided for the restoration of the auxiliary information. However, the amount of these information is really small which may not affect the applications. For example, when the embedding secret data are 100000 bits, the units which have the different grayscales as cover image are only 18.

## IV. PERFORMANCE EVALUATION

In this section, experimental results of the proposed method are presented. As shown in Fig. 14, the embedding performance is evaluated by four test images with size 512×512 in color version, including Lena, Airplane, Lake, Baboon (which are downloaded from USC-SIPI, http://sipi.usc.edu/database). Here, we apply the peak signal-to-noise ratio (PSNR) to indicate the distortion between marked and original images.

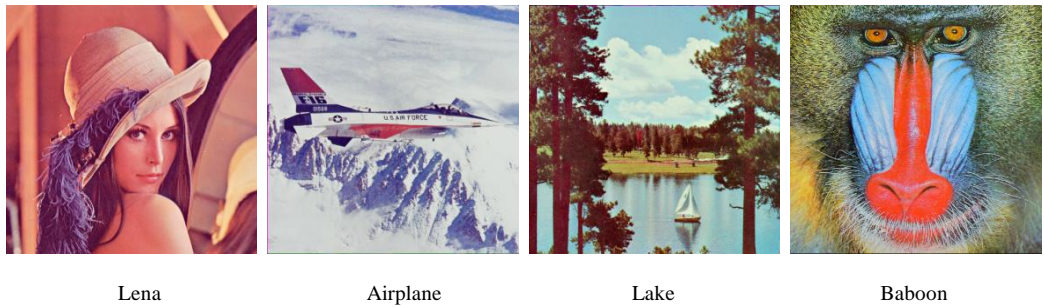

      Lena                    Airplane                   Lake                   Baboon

Fig. 14. Four test images.

Fig. 15 shows the Distortion-Capacity performance comparisons between our proposed method and [1], from which we could see that our proposed scheme outperforms Hou et al.'s approach on all the test images. Table I and Table II give the experimental data when the embedding capacities are 50000 and 150000 bits, respectively. On average, our method could gain 1.17 dB and 0.89 dB improvement over [1]. However, when trying to embed secret data into Baboon, the gaining performance is not large. It is because Baboon has abundant details and is difficult for predictor to obtain accurate results.

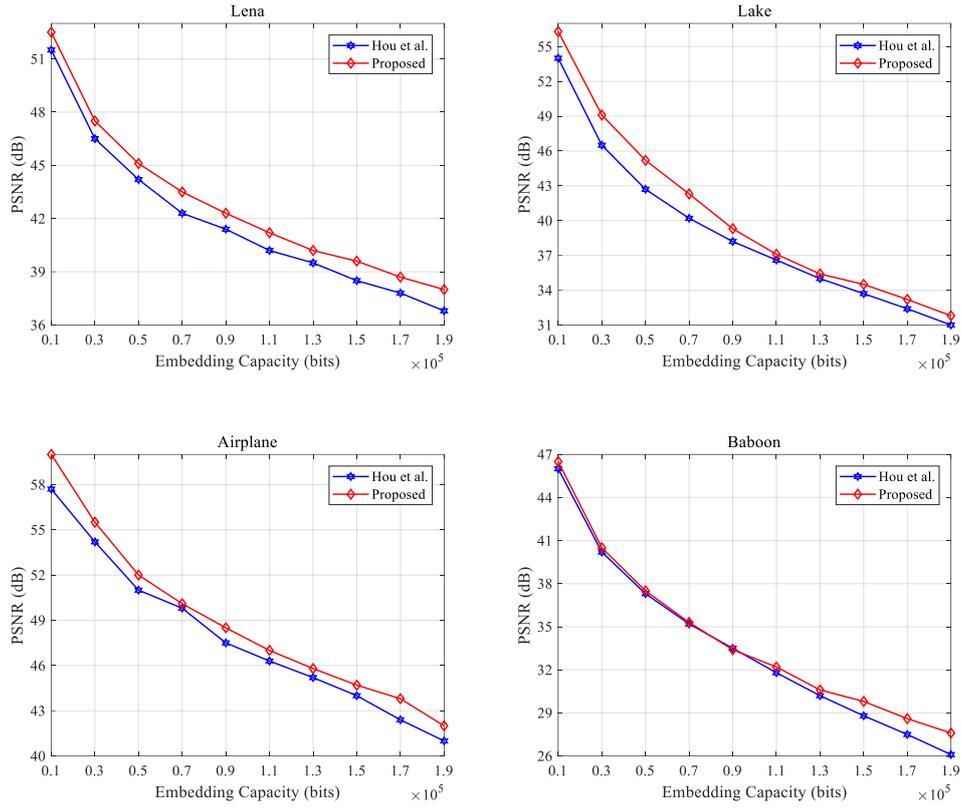

Fig. 15 Performance comparison on four test images.

TABLE I

PSNR(dB) COMPARISON FOR EMBEDDING CAPACITY OF 50000 BITS

| IMAGE | LENA | AIRPLANE | LAKE | BABOON | AVERAGE |
|---|---|---|---|---|---|
| HOU et al. | 44.18 | 51.06 | 42.86 | 37.29 | 43.80 |
| PROPOSED | **45.10** | **52.05** | **45.26** | **37.47** | **44.97** |

TABLE II

PSNR(dB) COMPARISON FOR EMBEDDING CAPACITY OF 150000 BITS

| IMAGE | LENA | AIRPLANE | LAKE | BABOON | AVERAGE |
|---|---|---|---|---|---|
| HOU et al. | 38.53 | 43.91 | 33.71 | 28.83 | 36.25 |
| PROPOSED | **39.37** | **44.73** | **34.55** | **29.79** | **37.14** |

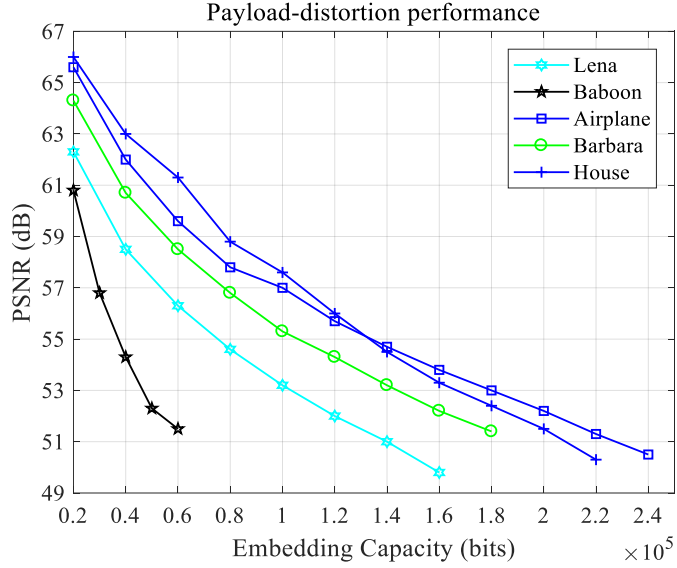

Fig. 16 Distortion-capacity performances on images of Yao et al.'s method [22].

For traditional RDH [19–22] methods on color images, channel relations are often exploited for obtaining better predicting results, by which the total distortion is reduced. Among all these methods, guided filtering based RDH (GF-RDH) has better performance than the other state-of-the-art, color image RDH methods, in which a linear transform model from reference channels to the current channel is established to improve the predicting accuracy. In Fig. 16, we give the distortion-capacity performances of [22], from which we could see that marked images' qualities of GF-RDH are far better than those of our method. However, as presented in [1], these kinds of methods could not preserve the features of marked image, which means that the marked image could not be utilized for other feature based assignments.

## V. CONCLUSION

To enhance the embedding performance of [1], an adaptive embedding pattern and a two-level predictor are designed in this paper. For pixels that located in smooth region, two bits of secret data are embedded into the R channel to reduce the unit embedding distortion. Instead of using one prediction value for two bits embedding, two-level predictor is designed by re-predicting value of marked pixel after the first embedding. Experimental results show the effectiveness of our proposed method.

## VI. ACKNOWLEDGEMENTS

This work is supported in part by the Open Project Program of the National Laboratory of Pattern Recognition (NLPR) (Grant No. 201800030).